\documentclass[11pt]{article}

\usepackage{latexsym,amsmath}

\DeclareFontFamily{OT1}{rsfs10}{}
\DeclareFontShape{OT1}{rsfs10}{m}{n}{ <-> rsfs10 }{}
\DeclareMathAlphabet{\mathscript}{OT1}{rsfs10}{m}{n}

\newcommand{\nn}{\nonumber}
\newcommand{\ns}{\normalsize}

\newcommand{\tr}{\textrm{tr}}

\newcommand{\bea}{\begin{eqnarray}}
\newcommand{\eea}{\end{eqnarray}}

\def\a{\alpha}
\def\b{\beta}

\def\d{\delta}
\def\e{\epsilon}

\def\z{\psi}
\def\k{\kappa}
\def\l{\lambda}

\def\n{\nu}
\def\o{\omega}
\def\p{\pi}

\def\r{\rho}
\def\s{\sigma}

\def\z{\zeta}

\def\G{\Gamma}

\def\cC{{\cal C}}

\def\cN{{\cal N}}

\def\Ib{\bar{I}}
\def\Jb{\bar{J}}
\def\Kb{\bar{K}}
\def\Lb{\bar{L}}

\begin{document}


\begin{titlepage}

\vspace{-5cm}

\title{
   \hfill{\ns OUTP-99-37P, UPR-854T} \\[3em]
   {\LARGE Symmetric Vacua in Heterotic M--Theory}\\[1em] } 
\author{
   {\ns Andr\'e Lukas$^1$ and Burt A.~Ovrut$^2$}\\[0.5em]
   {\it\ns $^1$Department of Physics, Theoretical Physics, 
       University of Oxford} \\[-0.5em]
   {\it\ns 1 Keble Road, Oxford OX1 3NP, United Kingdom}\\ 
   {\it\ns $^2$Department of Physics, University of Pennsylvania} \\[-0.5em]
   {\it\ns Philadelphia, PA 19104--6396, USA}\\} 
\date{}

\maketitle

\begin{abstract}
Symmetric vacua of heterotic M--theory, characterized by vanishing
cohomology classes of individual
sources in the three--form Bianchi identity, are analyzed on smooth Calabi--Yau 
three--folds. We show that such vacua do not exist for elliptically
fibered Calabi--Yau spaces. However, explicit examples are found
for Calabi--Yau three--folds arising as intersections in both unweighted 
and weighted projective space. We show that such
symmetric vacua can be combined with attractive phenomenological
features such as three generations of quarks and leptons. 
Properties of the low energy effective actions associated with symmetric 
vacua are discussed. In particular, the gauge kinetic functions receive 
no perturbative threshold corrections, there are no corrections to the 
matter field K\"ahler metric and the associated five--dimensional effective 
theory admits flat space as its vacuum. 
\end{abstract}

\thispagestyle{empty}

\end{titlepage}


\section{Introduction:}

A general property of theories containing branes is that bulk fields
are excited by sources located on those branes. This effect is
particularly pronounced in heterotic M--theory~\cite{HW1,HW2,W,H}
where $\cN=1$ supersymmetric $E_8$ gauge multiplets are localized on 
two ten--dimensional
orbifold planes in eleven--dimensional space--time. These boundary sources
lead, roughly, to a linear variation of the bulk fields along the
single transverse direction and, hence, to strong effects that grow
with the size of this dimension.

When constructing vacua of heterotic M--theory preserving four
supercharges, these effects cause a deformation of the zeroth
order Calabi--Yau background~\cite{W}. This deformation can be
determined using a strong coupling expansion and is, at
present, known only in the linearized approximation~\cite{W,low}.
Requiring the validity of this approximation places a bound
on the size of the eleventh dimension~\cite{bd} and, hence, certain regions
of the moduli space are unavailable.  In the
four--dimensional effective action, these deformations
induce threshold corrections to
the gauge kinetic functions~\cite{W,bd,hp1,low} which differ on the observable
and hidden sector branes and can lead to
a strong gauge coupling in one of the sectors. Furthermore, they
induce corrections to the matter field K\"ahler potential~\cite{low}. 

Such deformations of the Calabi--Yau background occur quite
generically in heterotic M--theory. They are caused by source
terms in the Bianchi identity of the M--theory three--form
field. These source terms have support on each of the orbifold
planes and, in general, are non--vanishing cohomologically. 
As an example, the standard embedding of the
spin connection into one of the $E_8$ gauge groups leads to
cohomologically non--trivial sources on each of the orbifold planes.
Essentially, this happens
because the gravitational contribution to the Bianchi identity is
split equally between the two orbifold planes. Generically, the same property is
shared by non--standard embedding vacua~\cite{stieb,benakli,lpt,nse} as well.
Typically, the sources turn out to be non--trivial in cohomology
and the Calabi--Yau background receives corrections. However, we will show
that not all non--standard embedding vacua have this property.

\vspace{0.4cm}

Specifically, in this paper, we would like to study the 
possibility of setting the
sources in the three--form Bianchi identity to zero individually, on each
orbifold plane, at least in
cohomology. Vacua with this property are called ``symmetric vacua''.
We want to emphasize that the concepts of symmetric vacua and standard
embeddings are different, the latter always having non--vanishing individual
sources in the three--form Bianchi identity. For
heterotic M--theory on K3, symmetric vacua have been studied in ref.~\cite{dmw}.
In ref.~\cite{stieb}, it has been shown that a class of
heterotic orbifold models with vanishing threshold corrections
exists. In these models, the vanishing corrections are 
due to a combination of topological
properties and special properties at the orbifold point. 

Here, we would like to construct symmetric vacua of heterotic
M--theory based on smooth Calabi--Yau three--folds. From what we have said,
this requires non--standard embedding vacua and, hence, the analysis
of certain classes of semi--stable holomorphic vector bundles $V$ over 
Calabi--Yau three--folds
$X$. As we will see, the relevant vector bundles 
needed to construct symmetric vacua are those with the property that
$c_2(V)=\frac{1}{2}c_2(TX)$, where $c_{2}(V)$ and $c_{2}(TX)$ are the second
Chern classes of $V$ and the tangent bundle of $X$ respectively. 
We will call such vector bundles
symmetric as well. Symmetric vacua have interesting properties, such
as the absence of the threshold and K\"ahler potential corrections to the 
low energy effective action mentioned above. They also admit flat space as
the vacuum solution to the associated five--dimensional effective theory.
It is also possible that they constitute 
valid solutions even in the region of moduli space with large orbifold 
radius, which is usually not accessible.

\vspace{0.4cm}

In section 2, we start by reviewing the general context and
by presenting some of the essential formulae. Section 3 is devoted to the
study of elliptically fibered Calabi--Yau spaces with sections and
semi--stable holomorphic bundles of the Friedman--Morgan--Witten type~\cite{FMW}. 
In section 4, we move on to discuss Calabi--Yau spaces defined as intersections in
unweighted and weighted projective spaces and holomorphic vector bundles of
the monad type. Finally, in section 5, we discuss the special 
properties of symmetric vacua and their associated low energy effective actions.

Our results can be summarized as follows. We show that, within the
class of elliptically fibered Calabi--Yau three--folds and Friedman--Morgan--Witten
vector bundles, no symmetric vector bundles and, hence, no symmetric vacua
exist. This confirms the expectation that symmetric vacua are, in
fact, rare and that generically Calabi--Yau vacua do receive corrections.
For Calabi--Yau three--folds defined as intersections in projective space,
we first prove several general properties of symmetric bundles (of
the monad type) such as a lower bound on the number of generations. Then, we
construct explicit examples of symmetric bundles on various
Calabi--Yau spaces within this class. Again, such bundles turn out
to be relatively rare. For example, for the five Calabi--Yau three--folds
defined as intersections in an unweighted projective space, there
exist exactly four symmetric bundles (of the monad type), three for
the quintic polynomial and one for the intersection of two cubic 
polynomials in ${\bf CP}^5$. Furthermore, using two of our examples, 
one in unweighted and
the other in weighted projective space, we show that symmetric vacua
can be combined with phenomenologically interesting properties
such as three generations of quarks and leptons. To first non--trivial order, 
symmetric vacua do not receive corrections at the level of the Calabi--Yau
zero modes. However, massive first order corrections are generically
present. We point out that, due to the
vanishing massless vacuum corrections, all first order strong coupling corrections
to the associated four-- and five--dimensional low energy 
effective actions vanish for symmetric vacua. In particular, the 
threshold corrections to the
gauge kinetic functions and the correction to the matter field
K\"ahler metric vanish. We also speculate that symmetric vacua might
be valid in the region of moduli space with large orbifold size
where the strong coupling expansion breaks down in the non--symmetric
case.

\section{General Framework:} 

In this section, we would like to present the general framework for
the discussion. Our starting point is the effective action for the strongly
coupled heterotic string~\cite{HW1,HW2} given by M--theory on
the orbifold $S^1/Z_2$. As usual, the orbifold coordinate
$x^{11}$ is taken to be in the range $x^{11}\in [-\p\r ,\p\r ]$ with
the $Z_2$ symmetry acting as $x^{11}\rightarrow -x^{11}$. This leads
to two fixed ten-planes at $x^{11}=0$ and $\p\r$, each of which carries an
$\cN =1$ supersymmetric $E_8$ gauge multiplet.

We would like to consider vacua of this theory associated with $\cN =1$
supergravity theories in four dimensions. To lowest order, this
implies the space--time structure
\begin{equation}
   M_{11}=M_{4} \times S^{1}/Z_{2} \times X
\label{M11}
\end{equation}
where $M_4$ is four--dimensional Minkowski space and $X$ is a
Calabi--Yau three--fold. As is well--known~\cite{W,H}, the above direct
product structure is valid only to lowest order in an expansion in the
eleven--dimensional Newton constant $\k$. More precisely, the first corrections
appear at order $\k^{2/3}$ and produce a ``deformation'' of
the above space in both the orbifold and Calabi--Yau
components. The key ingredient in the theory that leads to these
deformations is the Bianchi identity~\footnote{Indices
$\Ib ,\Jb ,\Kb ,\dots = 0,\dots 9$ specify the 10--dimensional space
orthogonal to the orbifold.}
\bea
 (dG)_{11\Ib\Jb\Kb\Lb} &=& 4\sqrt{2}\p\left(\frac{\k}{4\p}
                           \right)^{2/3}\left(J^{(0)}\d (x^{11})+J^{(N+1)}
                           \d (x^{11}-\p\r )+\right.\nn \\
                       &&\left.\qquad\qquad\qquad\qquad\frac{1}{2}
                         \sum_{i=1}^NJ^{(i)}(\d (x^{11}-x_n)+\d (x^{11}+x_n))
                           \right)_{\Ib\Jb\Kb\Lb}\; .\label{G}
\eea
for the field strength $G=6\, dC$ of the M-theory three--form
field $C$. Here $J^{(0)}$ and $J^{(N+1)}$ are sources supported on the orbifold
planes and defined in terms of the two $E_8$ field strengths $F^{(1)}$,
$F^{(2)}$ and the curvature $R$ by 
\begin{equation}
\begin{aligned}
 J^{(0)} &= -\frac{1}{16\p^2}\left.\left(\tr F^{(1)}\wedge F^{(1)} 
      - \frac{1}{2}\tr R\wedge R\right)\right|_{x^{11}=0} \; , \\
 J^{(N+1)} &= -\frac{1}{16\p^2}\left.\left(\tr F^{(2)}\wedge F^{(2)} 
      - \frac{1}{2}\tr R\wedge R\right)\right|_{x^{11}=\p\r} \; .
\end{aligned}
\label{J} 
\end{equation}
For the vacua under consideration, the gauge fields $F^{(1)}$ and
$F^{(2)}$ are associated with holomorphic vector bundles $V_1$ and $V_2$
on the Calabi--Yau space $X$ while the curvature $R$ is
associated with the tangent bundle $TX$. 
Note that the gravitational contribution $\tr(R \wedge R)$ to the Bianchi
identity has been split equally between the two orbifold planes. This can be
seen from the factors $1/2$ in front of $\tr(R \wedge R)$ that appear in
eq.~\eqref{J}. 
For generality, we have also
allowed for additional sources $J^{(i)}$, where $i=1,\dots ,N$, in the
Bianchi identity which could arise from five--branes in the vacuum. Such
vacua with five--branes have been constructed and analyzed in
ref.~\cite{nse,dlow1,dlow2}. Here, we will only need some elementary
properties. The $N$ five--branes are oriented transversely to the
orbifold and located at $x^{11}=x_1,\dots ,x_N$. To preserve
four--dimensional Poincar\'e invariance and $\cN =1$ supersymmetry,
they stretch across the uncompactified space $M_4$ and wrap on
holomorphic curves $\cC_2^{(i)}$ within the Calabi--Yau space $X$. This
orientation implies for their sources that
\begin{equation}
   J^{(i)} = \d(\cC_2^{(i)}) \; ,\qquad i=1,\dots ,N\; ,
\label{Jdef}
\end{equation}
where $\d(\cC_2^{(i)})$ is the Poincar\'e dual four--form to the holomorphic
curve $\cC_2^{(i)}$.

Integrating Bianchi identity~\eqref{G} over an arbitrary four--cycle in the
Calabi--Yau space times the orbifold cycle, one finds~\footnote{We will
consider vector bundles $V$ with $c_1(V)=0$ throughout the paper.}
\begin{equation}
 c_2(V_1)+c_2(V_2)+[W]=c_2(TX) \label{coh}
\end{equation}
where 
\begin{equation}
 c_2(V_i)=-\frac{1}{16\pi^{2}}\left[ \tr F^{(i)}\wedge F^{(i)}\right]\; , \qquad
 c_2(TX)= -\frac{1}{16\pi^{2}}\left[ \tr R \wedge R \right] \;
\label{eq:burt1}
\end{equation}
are the second Chern classes for the vector bundle $V_i$ and the tangent bundle
$TX$ of $X$ respectively
and 
\begin{equation}
[W]= \sum_{i=1}^N\left[ \d(\cC_2^{(i)})\right]
\label{eq:burt2}
\end{equation}
is the four--form class of the five--branes. 
The brackets $\left[ \dots \right]$ indicate the cohomology class of the 
associated four--form.
This topological condition guarantees
anomaly--freedom of the vacuum and it is necessary
(and sufficient) for the Bianchi identity to be soluble. It
states that the total right hand side of the Bianchi identity is topologically
trivial, as it should be. However, this is not the case for 
the cohomology classes of the individual source terms $\left[ J^{(n)} \right]$, 
for $n=0, \dots,N+1$, which are generally non--vanishing. Hence, the sources 
themselves generally do not vanish, that is, $J^{(n)} \neq 0$. This
leads to a non--vanishing field strength $G$ which, in
turn, causes a deformation of the space--time~\eqref{M11}. 

\vspace{0.4cm}

As a familiar example, let us consider the standard embedding. 
The anomaly constraint \eqref{coh} can be satisfied if one takes
\begin{equation}
c_2(V_1)=c_2(TX), \qquad  c_2(V_2)=0
\label{eq:hello1}
\end{equation}
and
\begin{equation}
[W]=0.
\label{eq:hello2}
\end{equation} 
The last statement asserts that the standard embedding does not 
allow five--branes in the vacuum, as expected. Therefore, $N=0$. 
Note from eq.~\eqref{J} and \eqref{eq:hello1} that the cohomology classes of 
the remaining sources are given by 
\begin{equation}
 \left[ J^{(0)}\right] = -\frac{1}{32\p^2}\left[ \tr (R\wedge R) \right]\; ,\qquad
 \left[ J^{(1)}\right] = \frac{1}{32\p^2} \left[\tr (R\wedge R)\right]\; . 
\end{equation}
Since $c_2(TX)\neq 0$, the cohomology class of the source on each of the 
orbifold planes is non--vanishing. This is due to
the aforementioned fact that the gravitational part of the Bianchi 
identity has been equally distributed onto each of the two orbifold planes. 
Solution~\eqref{eq:hello1},~\eqref{eq:hello2} is an attractive choice, 
since it 
can be explicitly realized at the level of fields by
embedding the spin connection into the first $E_{8}$ gauge group and 
choosing the
second $E_{8}$ gauge vacuum to be trivial. That is
\begin{equation} 
\tr(F^{(1)}\wedge F^{(1)})= \tr(R \wedge R)\; , \qquad
\tr(F^{(2)}\wedge F^{(2)})=0\; . \label{se1}
\end{equation}
 It follows from eq.~\eqref{J} that the sources are explicitly given by
\begin{equation}
 J^{(0)} = -\frac{1}{32\p^2} \tr (R\wedge R)\; ,\qquad
 J^{(1)} = \frac{1}{32\p^2} \tr (R\wedge R)\; . 
\end{equation}
Since the source cohomology classes are non-vanishing, it follows 
that the sources themselves cannot be set to zero. 
Therefore, the standard embedding leads to a non--vanishing field strength
$G$ and, hence, to a deformation of the Calabi--Yau vacuum~\cite{W}.

\vspace{0.4cm}

The main point of this paper is to look for solutions of the anomaly 
constraint \eqref{coh} which possibly allow for the vanishing of the 
right hand side of the 
Bianchi identity. In other words, we are interested in vacua 
where each source term $J^{(n)}$, for $n=0, \dots,N+1$,
on the right hand side of the Bianchi identity potentially vanishes.
For such vacua, the four--form field strength $G$ can be set to zero
and, hence, the space--time~\eqref{M11} remains uncorrected, at least to first
non--trivial order in the $\kappa$ expansion. The task of finding such 
vacua can be broken into two steps. First, if each of the sources 
is to vanish, it necessarily must vanish in cohomology, $[J^{(n)}]=0$. 
This is easily achieved for the five--brane sources. We simply
consider vacua without five--branes, that is, we take  $N=0$. In this
case, the five--brane sources $J^{(i)}$ and their cohomology classes simply do
not exist. In the following, we concentrate exclusively on
vacua without five--branes. The situation is more complicated for
the sources on the orbifold planes. For those sources to be
cohomologically trivial, that is $[J^{(0)}]=[J^{(1)}]=0$, it follows from
eq.~\eqref{J} that we need to set
\begin{equation}
\left[\tr(F^{(1)}\wedge F^{(1)})\right]= \left[\tr(F^{(2)}\wedge F^{(2)})\right]
= \frac{1}{2} \left[\tr(R\wedge R)\right].
\label{eq:burt3}
\end{equation}
Consequently, we have
\begin{equation}
 c_2(V_1)=c_2(V_2)=\frac{1}{2}c_2(TX)\; .\label{symm}
\end{equation}
which satisfies the topological constraint~\eqref{coh} if we take 
\begin{equation}
[W]=0,
\label{eq:hello3}
\end{equation}
consistent with the above assumption.
We will refer to vacua with property~\eqref{symm} and no
five--branes as symmetric vacua. Second, having chosen a Calabi--Yau
three--fold $X$ and vector bundles $V_1$ and $V_2$ realizing such a symmetric
vacuum, one has removed the topological obstruction of setting the
orbifold sources to zero. To actually set the sources to zero, 
however, one needs to
explicitly choose a spin connection on $TX$ and specific gauge connections
on $V_1$ and $V_2$ so that $J^{(0)}=J^{(1)}=0$. 
This would imply that
\begin{equation}
\tr(F^{(1)}\wedge F^{(1)})= \tr(F^{(2)}\wedge F^{(2)})
= \frac{1}{2} \tr(R\wedge R).
\label{eq:burt100}
\end{equation}
Unlike in the case of the standard embedding where the
relations~\eqref{se1} are satisfied at the level of fields,
it is not clear that 
vacuum solutions satisfying constraint~\eqref{eq:burt100} can be 
achieved. Therefore,
for the purposes of this paper, we restrict ourselves to solving the
topological part of the problem specified by eq.~\eqref{eq:burt3}.
However, if we cannot be sure that sources $J^{(0)}$ and $J^{(1)}$ vanish,
then it is conceivable that the field strength $G$ is again
non--vanishing and,  hence, spacetime~\eqref{M11} is deformed.
What simplification, then, has been achieved by using symmetric vacua?
The answer is, considerable simplification. To see this, note
that we have shown elsewhere~\cite{low,losw2} that the corrections, 
due to the deformation of space--time~\eqref{M11}, to the the effective 
action of the zero modes are all proportional to charges $\beta_{i}$
defined by
\begin{equation}
\b_i(*\o^i)= c_2(V_1)-\frac{1}{2}c_2(TX)=-c_2(V_2)+\frac{1}{2}c_2(TX) 
\end{equation}
Here $\{\o_i \}_{i=1,\dots ,h^{1,1}}$ is a basis of $H^{1,1}(X)$.
The quantities on the right hand side of this equation are exactly
those that are set to zero for symmetric vacua. We conclude that
all charges $\b_i$ vanish for symmetric vacua. Therefore, for symmetric 
vacua, even if there is a deformation of space--time~\eqref{M11} 
due to non--vanishing sources $J^{(0)}$ and $J^{(1)}$, to first
non--trivial order, this deformation does not effect the low energy
zero mode action as a consequence of the vanishing of the 
cohomology classes $\left[J^{(0)}\right]=\left[J^{(1)}\right]=0$. This 
motivates us to search for symmetric vacua in this paper.
Finally, note that  eq.~\eqref{symm} is much stronger than the anomaly 
cancellation condition~\eqref{coh} so that we are, in fact, looking for a small
sub--class of all non--anomalous vacua.

The problem we are going to address, then, is to find,
for a given Calabi--Yau three--fold $X$, holomorphic vector bundles $V$ with
the property that
\begin{equation}
 c_2(V) = \frac{1}{2}c_2(TX)\; .\label{symm1}
\end{equation}
We will call such vector bundles symmetric bundles of the Calabi--Yau
space $X$. It is understood, in what follows, that a symmetric
embedding can then be constructed by choosing both relevant bundles $V_1$
and $V_2$ of the symmetric type. Note again that the standard embedding
implies that $c_2(V)=c_2(TX)$, which
differs from the condition~\eqref{symm1} by a crucial factor of $1/2$.
Hence, the standard embedding does not provide us with symmetric bundles.

For phenomenological reasons, we will be interested in the number of 
chiral generations of quarks and leptons
associated with a (symmetric) bundle $V$. It is given by
\begin{equation}
 N_{\rm gen} = \frac{1}{2}\left|\int_X c_3(V)\right|\; .\label{Ngen}
\end{equation}
Furthermore, we will focus on holomorphic vector bundles with structure group
$G=SU(n)$ only. This is phenomenologically motivated, since the
choices $n=3,4$ and $5$ lead to the low--energy groups (the commutants of $G$ in
$E_8$) $E_6$, $SO(10)$ and $SU(5)$, that is, to attractive grand--unified
groups.

\section{Elliptically Fibered Calabi--Yau Three--Folds:}

We will now consider the possibility of symmetric vacua for
elliptically fibered Calabi--Yau three--folds. Let us first review the
properties of these spaces that are essential to our discussion.

The relevant Calabi--Yau three--folds $X$ are given as an elliptic
fibration over a two--fold base $B$ with section $\s :B\rightarrow X$.
It can be shown~\cite{MV} that the base space $B$ is restricted to
del Pezzo, Enriques, or Hirzebruch surfaces or certain blow--ups of the
latter. The second Chern class of such a Calabi--Yau space can be
expressed in terms of the base $B$ and the section $\s$ as
\begin{equation}
 c_2(TX) = c_2(B)+11\, c_1(B)^2+12\, \s c_1(B)\; ,\label{c2TX}
\end{equation}
where $c_{1}(B)$ and $c_{2}(B)$ are the first and second Chern classes of the
base respectively. Next, we have to specify the class of holomorphic 
vector bundles over elliptically fibered Calabi--Yau spaces on which 
we intend to focus. We concentrate on the vector bundles with structure 
group $SU(n)$ constructed by Friedman, Morgan and Witten~\cite{FMW}. 
These bundles are parameterized by a class $\eta\in H_{2}(B)$ and a 
rational number $\l$ subject to the constraints
\begin{align}
   & \text{$n$ is odd},  \quad \l = m+\frac{1}{2} 
\label{codd} \\
   & \text{$n$ is even},  \quad 
      \l = m, \quad \eta = c_1(B) \!\!\mod 2 
\label{ceven}
\end{align}
where $m$ is an integer. The second Chern class of these bundles $V$ can
be expressed in terms of $\eta$, $\l$, $n$ and properties of the
fibration. One finds~\cite{FMW} that
\begin{equation}
 c_2(V) = \eta \s - \frac{1}{24} c_1(B)^2 \left(n^3 - n\right) 
              + \frac{1}{2} \left(\l^2 - \frac{1}{4}\right) n \eta 
                      \left(\eta - nc_1(B)\right) \label{c2V}\; .
\end{equation}

\vspace{0.4cm}

Let us now try to find symmetric vector bundles in this class. The
obstruction to having such a symmetric bundle is given by
\begin{equation}
 U = c_2(V)-\frac{1}{2}c_2(TX)\; .
\end{equation}
In order to solve the equation $U=0$, we split $U$ into a base component and a
fiber component as
\begin{equation}
 U=U_B+aF .
\end{equation}
Here $U_B$ is a second homology class in the base while $a$ is an
integer that counts the number of fibers $F$. Inserting the above Chern
classes, one finds
\bea
 U_B &=& \sigma\left( \eta - 6c_1(B)\right) \\
 a &=& -\frac{1}{24}c_1(B)^2(n^3-n)+\frac{1}{2}\left(\l^2-
       \frac{1}{4}\right)n\eta\left(\eta -nc_1(B)\right)-
       \frac{11}{2}c_1(B)^2-\frac{1}{2}c_2(B)\; .\label{a}
\eea
For a symmetric bundle, we must demand that $U_B=0$. This fixes the bundle
parameter $\eta$ to be
\begin{equation}
 \eta = 6c_1(B)\; .\label{eta}
\end{equation}
{}From eqs.~\eqref{codd} and~\eqref{ceven}, we see that this choice of
$\eta$ is always acceptable for odd $n$. For even $n$, it may or may
not be compatible  with the constraint \eqref{ceven}. Keeping
this restriction in mind, we try to satisfy the remaining condition
for a symmetric bundle, namely that $a=0$. Inserting $\eta$ in
eq.~\eqref{eta} into~\eqref{a} leads to the relation
\begin{equation}
 c_2(B) = N(\l ,n)\, c_1(B)^2 \label{c12}
\end{equation}
between the first and second Chern class of the base. The numbers
$N(\l ,n)$ are given by
\begin{equation}
 N(\l ,n) = \left( 6\l^2-\frac{3}{2}\right) n(6-n)
            -\frac{1}{12}(n^3-n)-11\; .
\end{equation}
What remains to be checked is whether the relation~\eqref{c12} can be
satisfied for certain base spaces $B$ and numbers $\l$, $n$.
To do this, let us look at the explicit values of $c_1(B)^2$ and
$c_2(B)$ for the various allowed base spaces $B$ as given in
Table~\ref{Btable}.
\begin{table}
 \begin{center}
  \begin{tabular}{|c|c|c|c|}
   \hline
    Name&$B$&$c_1(B)^2$&$c_2(B)$ \\ \hline\hline
    del Pezzo&$dP_r\; ,r=1,\dots ,9$&$9-r$&$3+r$ \\ \hline
    Hirzebruch&$F_r$&$8$&$4$ \\ \hline
    Enriques&$E$&$0$&$12$ \\ \hline 
   \end{tabular}\\[0.5cm] 
  \end{center}
  \caption{Base spaces and their Chern classes}
  \label{Btable}
\end{table}
For the Enriques surface $E$ and the del Pezzo
surface $dP_9$ we have 
\begin{equation}
c_1(B)^2=0 , \qquad c_2(B)\neq 0. 
\label{eq:hello5}
\end{equation}
This is
incompatible with the relation~\eqref{c12} and, hence,
no symmetric bundles exist for these base spaces. From
Table~\ref{Btable}, we have for all other base
spaces that 
\begin{equation}
\frac{1}{2}\leq\frac{c_2(B)}{c_1(B)^2}\leq 11 . 
\label{eq:hello6}
\end{equation}
On the
other hand, it is easy to show that the numbers $N(\l ,n)$ satisfy 
either $N(\l ,n)<0$ or $N(\l ,n)\geq 20$ for all allowed values
of $n$ and $\l$. Hence the relation~\eqref{c12} can never be
satisfied.

We conclude that for the base spaces listed in Table~\ref{Btable}, no
symmetric vector bundles of the Friedman--Morgan--Witten type exist.
This result shows clearly that the property of being a symmetric
bundle is quite restrictive.

\section{Intersections in Weighted Projective Space:}

The condition~\eqref{symm1} for a symmetric vector bundle is an
equation in the vector space $H^{2,2}(X)$ and, therefore, provides
$h^{1,1}$ constraints on the bundle data. In the previous section we
have, unsuccessfully, looked for symmetric bundles over elliptically
fibered Calabi--Yau three--folds. For those Calabi--Yau space, we always have
$h^{1,1}\geq 2$. Therefore, it seems worthwhile to minimize the
number of constraints and consider Calabi--Yau spaces with
$h^{1,1}=1$.

\vspace{0.4cm}

A large class of Calabi--Yau three--folds with this property is provided by
intersections in both unweighted and weighted projective space. We will
focus our attention on such spaces in this section~\footnote{Weighted 
projective space is
singular. If these singularities intersect the Calabi--Yau space they
have to be blown up in order to arrive at a smooth manifold.
This creates new classes and, hence, a manifold with $h^{1,1} >
1$. For the purpose of this paper, we concentrate on cases where this
does not happen, so that indeed $h^{1,1}=1$.}. Partially, this is
motivated by earlier approaches to heterotic model building within this
class, particularly in ref.~\cite{dg,kachru}.
Let us review the 
relevant properties of these spaces following ref.~\cite{cls,huebsch}.
The starting point is the weighted projective space ${\bf CP}^{N+3}_{\bf
w}$ with homogeneous coordinates $(X^\n )_{\n =0,\dots ,N+3}$ and
weights ${\bf w}=(w_\n )_{\n =0,\dots ,N+3}$. The space
$X\subset {\bf CP}^{N+3}_{\bf w}$ is defined as the zero locus of $N$
polynomials $P_1(X),\dots ,P_N(X)$ with degrees
${\bf p} = (p_\a )_{\a =1,\dots ,N}$, where $p_\a = \mbox{deg}(P_\a )$.
The first Chern class of $X$ is given by
\begin{equation}
 c_1(TX) = \left(\sum_{v=0}^{N+3}w_\n -\sum_{\a =1}^Np_\a\right) J 
\end{equation}
Here $J=c_1({\cal O }(1))$ is the first Chern class of the hyperplane
bundle ${\cal O}(1)$. For $X$ to be a Calabi--Yau space, we want that
$c_1(TX)=0$ and, hence, that
\begin{equation}
 \sum_{v=0}^{N+3}w_\n = \sum_{\a =1}^Np_\a \; .
\end{equation}
Provided this condition is satisfied, one finds for the second Chern
class
\begin{equation}
 c_2(TX) = \frac{1}{2}\left(\sum_{\a =1}^Np_\a^2-
           \sum_{\n =0}^{N+3}w_{\n}^{2}\right) J^2\; .
\end{equation}
Another useful relation which we need below in order to evaluate the
number of generations is
\begin{equation}
 \int_X J^3 = \frac{\Pi_{\a =1}^Np_a}{\Pi_{\n =0}^{N+3}w_\n}\; .
 \label{intJ3}
\end{equation}

Next, we have to specify a class of holomorphic vector bundles. In the following, we
are going to use the so called monads defined by the short exact
sequence~\cite{dg,dk,kachru}
\begin{equation}
 0\longrightarrow V\longrightarrow \bigoplus_{a=1}^{n+M}{\cal O}(n_a)
 \stackrel{\otimes Q_a^i(X)}{\longrightarrow}\bigoplus_{i=1}^M{\cal O}(m_i)
 \longrightarrow 0\; .\label{seq}
\end{equation}
Here, the map between the two sums of line bundles is defined by the
polynomials $Q_a^i(X)$ with degree $\mbox{deg}(Q_a^i)=m_i-n_a$.
Consequently, one requires that
\begin{equation}
 m_i>n_a>0 \label{bounds}
\end{equation}
for all $i=1,\dots ,M$ and all $a=1,\dots ,n+M$. The vector bundle $V$
is then specified by the two sets of integers $(m_i)_{i=1,\dots ,M}$
and $(n_a)_{a=1,\dots ,n+M}$. It is denoted by
\begin{equation}
 V=V(m_1,\dots ,m_M ;n_1,\dots ,n_{n+M})\; .
\end{equation}
For the first Chern classes of such a bundle $V$, one finds that
\bea
 c_1(V) = \left(\sum_{a=1}^{n+M}n_a-\sum_{i=1}^Mm_i\right) J 
\label{burt4}
\eea
The condition that $V$ be a semi--stable bundle and the fact that
$h^{1,1}=1$ imply $c_1(V)=0$. As a result, $V$ is a bundle with 
structure group $SU(n)$. Therefore, we require 
\begin{equation}
 \sum_{a=1}^{n+M}n_a=\sum_{i=1}^Mm_i\; .
\end{equation}
For vanishing first Chern class, the second and third Chern classes are given
by
\bea
c_2(V) &=& -\frac{1}{2}\left(\sum_{a=1}^{n+M}n_a^2-\sum_{i=1}^Mm_i^2
            \right) J^2 \\
 c_3(V) &=& \frac{1}{3}\left(\sum_{a=1}^{n+M}n_a^3-\sum_{i=1}^Mm_i^3
            \right) J^3\; .\label{c3V}
\eea
Putting together eq.~\eqref{Ngen}, \eqref{intJ3} and \eqref{c3V} one
obtains for the number of generations
\begin{equation}
 N_{\rm gen} = \frac{1}{6}\left| \sum_{a=1}^{n+M}n_a^3-
               \sum_{i=1}^Mm_i^3\right|\frac{\Pi_{\a
               =1}^Np_\a}{\Pi_{\n = 0}^{N+3}w_\n}\; .
\label{eq:burt5}
\end{equation}

\vspace{0.4cm}

Before we discuss explicit examples, it is useful to prove some
general properties of symmetric bundles in the setting described
above. Let us focus on a specific Calabi--Yau three--fold $X$ represented
by weights $(w_\n )$ and polynomials of degree $(p_\a )$. For this
space $X$, define the quantity
\begin{equation}
 Q = \frac{1}{2}\left(\sum_{\a =1}^Np_\a^2-\sum_{\n
      =0}^{N+3}w_\n^2\right)\; .
\end{equation}
Then we are interested in vector bundles $V=V(n_a;m_i)$ on $X$ with
the properties
\bea
 \sum_{i=1}^Mm_i &=& \sum_{a=1}^{n+M}n_a\equiv S \\
 \sum_{i=1}^Mm_i^2 &=& \sum_{a=1}^{n+M}n_a^2+Q\; .
\eea
The first condition is just the statement that $c_1(V)=0$. The second
one states that $c_2(V)=\frac{1}{2}c_2(TX)$ and, hence, that $V$ is a
symmetric bundle. Furthermore, we define the quantity $C$ by
\bea
 \sum_{i=1}^Mm_i^3 &=& \sum_{a=1}^{n+M}n_a^3+C\; .
\eea
Using the bound~\eqref{bounds}, it is then easy to prove the
inequalities
\begin{equation}
 C>2Q\geq 2S \label{ineq}
\end{equation}
which must hold for any symmetric monad vector bundle on $X$. Recall
here that $Q$ depends on Calabi--Yau data only while
$C$ and $S$ depend on the vector bundle. The first part
of this inequality can be used, together with eq.~\eqref{eq:burt5}, 
to find a lower bound on the number of
generations associated with symmetric vector bundles. This bound is
given by
\begin{equation}
 N_{\rm gen} > \frac{Q}{3}\frac{\Pi_{\a =1}^Np_\a}{\Pi_{\n = 0}
               ^{N+3}w_\n}\; .\label{Ngenb}
\end{equation}
Note that the right hand side depends on data of the Calabi--Yau three--fold only.
Therefore, the above bound must be satisfied for any symmetric (monad)
vector bundle on the given Calabi--Yau space $X$. 

So far the integers $(m_i)$ and $(n_a)$ defining the vector bundle are
not bounded from above. Similarly, the number $M$ of line bundles is
not bounded. It turns out, however, that symmetric bundles are
possible only if those numbers do not exceed certain maximal values.
This is summarized in the following two statements.
\newtheorem{s1}{Statement}
\begin{s1}
 If $n_a>n_{\rm max}$ for any $a$ or $m_i>m_{\rm max}$ for any $i$, where
 $n_{\rm max}=Q+1-n-M$ and $m_{\rm max}=Q+2-2M$, then $V=V(m_i;n_a)$
 is not a symmetric bundle.
\end{s1}
\newtheorem{s2}[s1]{Statement}
\begin{s2}
 If $M>M_{\rm max}$ where $M_{\rm max}=Q-n$, then $V=V(m_i;n_a)$ is not
 a symmetric bundle.
\end{s2}
These two statements can easily be proven using the
inequality~\eqref{ineq}. They are useful because, for a given
Calabi--Yau three--fold, they only leave a finite set of monad vector
bundles as candidates for symmetric bundles. Scanning this finite set,
we can then find all symmetric monad bundles for a given Calabi--Yau
space.

\vspace{0.4cm}

We would now like to apply the above results to a number of explicit
examples, thereby showing that symmetric vector bundles indeed exist.
As already mentioned, for phenomenological reasons, we are mainly
interested in $SU(n)$ bundles with $n=3,4,5$. In the following
examples, we will focus on these three cases. We start with the five
Calabi--Yau spaces that can be defined as intersections in a single
unweighted projective space.\\[0.5cm]
{\bf Example 1:} Quintic polynomial in ${\bf CP}^4$ \\
\begin{table}
 \begin{center}
  \begin{tabular}{|c|c|c|}
   \hline
    $(n,M)$&$V=V(m_i;n_a)$&$N_{\rm gen}$\\ \hline\hline
    $(3,2)$&$(3,3;2,1,1,1,1)$&$35$\\ \hline
    $(4,3)$&$(3,2,2;1,1,1,1,1,1,1)$&$30$\\ \hline
    $(5,5)$&$(2,2,2,2,2;1,1,1,1,1,1,1,1,1,1)$&$25$\\ \hline
   \end{tabular}\\[0.5cm] 
  \end{center}
  \caption{Symmetric bundles for quintic in ${\bf CP}^4$}
  \label{quintic}
\end{table}
Using the above notation, the quintic is specified by $N=1$,
${\bf w}=(1,1,1,1,1)$ and ${\bf p}=(5)$. This leads to
\begin{equation}
 Q = 10\; ,\qquad \int_X J^3 = 5\; .
\end{equation}
The maximal integers for which symmetric bundles are possible are then
given by
\begin{equation}
 n_{\rm max} = 11-n-M\; ,\qquad m_{\rm max} = 12-2M\; ,\qquad
 M_{\rm max} = 10-n\; .
\end{equation}
Scanning the region $n_a\leq n_{\rm max}$, $m_i\leq m_{\rm max}$,
$M\leq M_{\rm max}$ one finds three symmetric monad bundles.
They are given in Table~\ref{quintic}.
We conclude that, for the quintic, there is exactly one symmetric
monad bundle for each rank $n=3,4,5$.\\[0.5cm]
{\bf Example 2:} Intersection of two cubic polynomials in ${\bf CP}^5$ \\
\begin{table}[ht]
 \begin{center}
  \begin{tabular}{|c|c|c|}
   \hline
    $(n,M)$&$V=V(m_i;n_a)$&$N_{\rm gen}$\\ \hline\hline
    $(3,3)$&$(2,2,2;1,1,1,1,1,1)$&$27$\\ \hline
   \end{tabular}\\[0.5cm] 
  \end{center}
  \caption{Symmetric bundles for the intersection of two
           cubics in ${\bf CP}^5$}
  \label{cubics}
\end{table}
This space is defined by $N=2$, ${\bf w}=(1,1,1,1,1,1)$ and
${\bf p}=(3,3)$. One finds
\begin{equation}
 Q=6\; ,\qquad \int_XJ^3=9
\end{equation}
and
\begin{equation}
  n_{\rm max} = 7-n-M\; ,\qquad m_{\rm max} = 8-2M\; ,\qquad
 M_{\rm max} = 6-n\; .
\end{equation}
Using these maximal numbers, one can show that there exists a unique
rank 3 symmetric vector bundle for this space. It is given in
Table~\ref{cubics}.\\[0.5cm]
{\bf Example 3:} Other intersections in unweighted projective space\\
There are three more Calabi--Yau spaces that can be defined as
intersections in a single unweighted projective space, namely the
intersection of a quadric and a quartic in ${\bf CP}^5$, the
intersection of two quadrics and a cubic in ${\bf CP}^6$ and the
intersection of four quadrics in ${\bf CP}^7$. Using the method
described above, one can show that no symmetric monad bundles of rank
$n=3,4,5$ exist for these three spaces.\\

To summarize our results so far, we have shown that for the five
Calabi--Yau spaces defined in a single unweighted projective space,
there exist exactly four symmetric monad bundles of rank $n=3,4,5$,
three for the quintic in ${\bf CP}^4$ and one for the intersection of two cubics in
${\bf CP}^5$. Let us now turn to two further examples in weighted
projective space.\\[0.5cm]
{\bf Example 4:} Degree 6 polynomial in ${\bf CP}^4_{1,1,1,1,2}$\\
\begin{table}[ht]
 \begin{center}
  \begin{tabular}{|c|c|c|}
   \hline
    $(n,M)$&$V=V(m_i;n_a)$&$N_{\rm gen}$\\ \hline\hline
    $(3,2)$&$(4,3;2,2,1,1,1)$&$36$\\ \hline
    $(3,4)$&$(3,3,3,3;2,2,2,2,2,1,1)$&$33$\\ \hline
    $(4,2)$&$(4,2;1,1,1,1,1,1)$&$33$\\ \hline
    $(4,3)$&$(3,3,3;2,2,1,1,1,1,1)$&$30$\\ \hline
    $(5,3)$&$(3,3,2;1,1,1,1,1,1,1,1)$&$27$\\ \hline
   \end{tabular}\\[0.5cm] 
  \end{center}
  \caption{Symmetric bundles for a Calabi--Yau space defined by a degree 6
           polynomial in ${\bf CP}^4_{1,1,1,1,2}$}
  \label{weighted1}
\end{table}
This space is characterized by $N=1$, ${\bf w}=(1,1,1,1,2)$ and
${\bf p}=(6)$. We find
\begin{equation}
 Q=14\; ,\qquad \int_XJ^3=3
\end{equation}
and
\begin{equation}
   n_{\rm max} = 15-n-M\; ,\qquad m_{\rm max} = 16-2M\; ,\qquad
 M_{\rm max} = 14-n\; .
\end{equation} 
Given those bounds, we find exactly the five symmetric bundles listed in
Table~\ref{weighted1}.\\[0.5cm]
{\bf Example 5:} Two degree 6 polynomials in ${\bf CP}^5_{1,1,2,2,3,3}$\\
This space is defined by $N=2$, ${\bf w}=(1,1,2,2,3,3)$ and ${\bf
p}=(6,6)$. It has also been used in ref.~\cite{kachru} to
construct a (non--symmetric) three--family model with the standard model gauge
group. We find
\begin{equation}
 Q=22\; ,\qquad \int_XJ^3=1
\end{equation}
and
\begin{equation}
 n_{\rm max} = 23-n-M\; ,\qquad m_{\rm max} = 24-2M\; ,\qquad
 M_{\rm max} = 22-n\; .
\end{equation}
Using those bounds, we find a total of 15 symmetric bundles for this
space. They are listed in Table~\ref{weighted2}.\\
\begin{table}[ht]
 \begin{center}
  \begin{tabular}{|c|c|c|}
   \hline
    $(n,M)$&$V=V(m_i;n_a)$&$N_{\rm gen}$\\ \hline\hline
    $(3,2)$&$(5,4;3,2,2,1,1)$&$24$\\ \hline
    $(3,2)$&$(5,5;4,3,1,1,1)$&$26$\\ \hline
    $(3,3)$&$(5,3,3;2,2,2,2,2,1)$&$23$\\ \hline
    $(3,3)$&$(4,4,4;3,2,2,2,2,1)$&$22$\\ \hline
    $(3,4)$&$(4,4,3,3;2,2,2,2,2,2,2)$&$21$\\ \hline
    $(3,4)$&$(4,4,4,4;3,3,3,3,2,1,1)$&$23$\\ \hline
    $(4,2)$&$(5,3;2,2,1,1,1,1)$&$22$\\ \hline
    $(4,3)$&$(4,4,3;2,2,2,2,1,1,1)$&$20$\\ \hline
    $(4,3)$&$(4,4,4;3,3,2,1,1,1,1)$&$21$\\ \hline
    $(4,5)$&$(4,3,3,3,3;2,2,2,2,2,2,2,1,1)$&$19$\\ \hline
    $(4,7)$&$(3,3,3,3,3,3,3;2,2,2,2,2,2,2,2,2,2,1)$&$18$\\ \hline
    $(5,2)$&$(5,2;1,1,1,1,1,1,1)$&$21$\\ \hline
    $(5,2)$&$(4,4;2,1,1,1,1,1,1)$&$19$\\ \hline
    $(5,4)$&$(4,3,3,3;2,2,2,2,1,1,1,1,1)$&$18$\\ \hline
    $(5,6)$&$(3,3,3,3,3,3;2,2,2,2,2,2,2,1,1,1,1)$&$17$\\ \hline
   \end{tabular}\\[0.5cm] 
  \end{center}
  \caption{Symmetric bundles for a Calabi--Yau space defined by two degree 6
           polynomials in ${\bf CP}^5_{1,1,2,2,3,3}$}
  \label{weighted2}
\end{table}

This concludes our list of explicit examples. We have seen that
symmetric vector bundles on Calabi--Yau three--folds in both unweighted and
weighted projective spaces exist. The five Calabi--Yau spaces in unweighted
projective space allow for exactly four symmetric (monad)
bundles of rank $n=3,4,5$. Our final two examples showed that it is
somewhat easier to find symmetric bundles on Calabi--Yau three--folds in
weighted projective space. Still, it is clear that such symmetric
bundles are relatively rare objects.

\vspace{0.4cm}

Based on the above experience, we would now like to ask whether
symmetric vacua can have interesting phenomenological properties, such
as three chiral families of quarks and leptons in the observable sector. 
The above list of
examples does not contain a single case with three generations. Given
the lower bound~\eqref{Ngenb} on the number of generations for
symmetric (monad) bundles, this is not surprising. However, as usual,
we are not necessarily interested in getting three generations on
the original Calabi--Yau three--fold $X$. Instead, in order to be able to
break the grand unified group by Wilson lines, we would like to
consider non--simply connected Calabi--Yau three--folds defined by $Y=X/D$,
where $D$ is a freely acting discrete automorphism group on
$X$. Assuming that we are able to lift the automorphism $D$ to the vector
bundle $V$ (so it defines a bundle $V_{Y}$ on $Y$), the ``new'' number of
generations on $Y$ is given by 
\bea
 N_{\rm gen}(Y)=N_{\rm gen}/|D|
\label{eq:burt6}
\eea
where $N_{\rm gen}$ is the number of generations on $X$ and
$|D|$ is the order of the group $D$. 
It is for this new number of generations that we require
\begin{equation}
N_{\rm gen}(Y)=3 .
\label{eq:hello7}
\end{equation}
Note that if $q:X \rightarrow Y$ is the covering map, then
\begin{equation}
c_{2}(V_{Y})=\frac{1}{|D|}q_{*}c_{2}(V), \qquad 
c_{2}(TY)=\frac{1}{|D|}q_{*}c_{2}(TX)
\label{last}
\end{equation}
and symmetry property~\eqref{symm} continues to hold on $Y$. Hence, the 
quotient vacuum is a symmetric vacuum.
In order to construct three--family quotient manifolds $Y$, 
the interesting symmetric vacua on $X$ are those with a generation number that 
is a multiple of three. Indeed, there are a few such examples contained 
in the above tables. 

There is one more constraint that has to be satisfied in order 
for the quotient symmetric vacuum to be consistent. This is the level matching
condition of ref.~\cite{vafa}. Let us briefly summarize this
constraint. Consider a Calabi--Yau three--fold defined as the intersection of
polynomials in projective space with coordinates $X^\n$. 
Let $D={\bf Z}_N$ be a discrete group with generator $g$
which acts on the coordinates as
\begin{equation}
 g:X^\n\rightarrow\a^{k_\n}X^\n\; ,
\label{nabob}
\end{equation}
where $\a=\exp(2\p i/N)$ and $k_\n$ are integer charges and assume that 
$D$ is an automorphism of this Calabi--Yau space.
Also assume that this automorphism lifts to a vector bundle $V$ over 
the Calabi--Yau 
three--fold. This will be the case if one chooses the ${\bf Z}_N$
charges $\tilde{k}_{a}$ of the coordinates $\z_a$ of the line bundles 
${\cal O}(n_a)$ that appear in the exact sequence~\eqref{seq} in a 
specific way~\cite{dk,kachru}, to be discussed below. 
For the two vector bundles $V_{1}$ and 
$V_{2}$ on the orbifold planes there are two sets of charges 
$\tilde{k}_{1a}$ and $\tilde{k}_{2a}$, 
respectively. Then, the level matching condition of ref.~\cite{vafa} 
states that these charges should satisfy
\bea
 \sum_\n k_\n^2
   &=&\sum_a\tilde{k}_{1a}^2+\sum_a\tilde{k}_{2a}^2\quad\mbox{mod}\quad 2N 
   \label{lm1}\\
 \sum_\n k_\n &=& \sum_a\tilde{k}_{1a} = \sum_a\tilde{k}_{2a} = 0
  \quad\mbox{mod}\quad 2 \label{lm2}
\eea
for $N$ even. For $N$ odd we only have the first constraint with $2N$
replaced by $N$. 

\vspace{0.4cm}

In order to make the above line of thought explicit, we consider the
Calabi--Yau three--fold of Example 2 defined by the intersection of two cubic
polynomials in ${\bf CP}^5$. As we have seen, this space has a unique 
rank 3 symmetric vector bundle
\begin{equation}
 V=V(2,2,2;1,1,1,1,1,1)
\label{again}
\end{equation}
with $N_{\rm gen}=27$. Let us choose the two cubic
polynomials
\begin{equation}
 P_1(X)=\sum_{\n =0}^5a_\n (X^\n )^3\; ,\qquad
 P_2(X)=\sum_{\n =0}^5b_\n (X^\n )^3\; .
\end{equation} 
to define the Calabi--Yau three--fold $X$. Then, for generic choices of the
coefficients $a_\n$ and $b_\n$, the manifold $X$ is non--singular. 
Furthermore, $X$ admits an
automorphism $D={\bf Z}_3\times{\bf Z}_3$ with generators $g_1$ and
$g_2$ acting as
\begin{equation}
 g_1:X^\n\rightarrow\a^{k_\n}X^\n\; ,\qquad
 g_2:X^\n\rightarrow\a^{l_\n}X^\n\
\label{nab1}
\end{equation}
where $\a=\exp(2\p i/3)$ and $k_\n$, $l_\n$ are integer charges. Pick,
for example
\begin{equation}
 {\bf k}=(0,0,1,1,2,2)\; ,\qquad {\bf l}=(0,1,2,0,1,2)\; .
\end{equation}
The set of fixed points under the ${\bf Z}_3\times{\bf Z}_3$ transformations
\eqref{nab1} can be shown to have complex dimension one. Generically, 
a three--fold $X$ in ${\bf CP}^5$ does not intersect a curve. Hence,
generically, $D$ is freely acting on $X$ and we can define the quotient
Calabi--Yau three--fold $Y=X/D$. We should also lift $D$ to an 
automorphism of the bundle $V$. As a first step, we have to choose
the ${\bf Z}_3\times{\bf Z}_3$
charges $(\tilde{k}_a,\tilde{l}_a)$ for the coordinates $\z_a$ of the
line bundles ${\cal O}(n_a)$ that appear in the exact
sequence~\eqref{seq}. Furthermore, we should pick explicit polynomials
$Q_a^i(X)$ in this exact sequence. These polynomials each inherit a  
${\bf Z}_3\times{\bf Z}_3$ charge from the charges on $X^{\nu}$. 
Then $D$ lifts to a symmetry of the vector bundle $V$
if the above charges 
are chosen in such a way that $\z_a Q_a^i(X)$ is invariant
under $D$ for all $a$ and $i$~\cite{dk,kachru}. For the case at hand, this can indeed
be done. For example, we can choose the charges of
$\z_a$ as
\begin{equation} 
\tilde{\bf k}=(0,1,1,2,2,2)\; ,\qquad \tilde{\bf l}=(0,1,2,1,2,2)\; .
\end{equation}
It follows from the structure of the vector bundle $V$~\eqref{again}, 
that all polynomials $Q_a^i(X)$ should be linear in $X^\n$.
It is not hard to show that these linear polynomials can be chosen so that 
they inherit the ${\bf Z}_3\times{\bf Z}_3$ charges
$(-\tilde{k}_a,-\tilde{l}_a)$ from the coordinates $X^\n$.
It is then clear that $\z_a Q_a^i(X)$ is indeed invariant for all $a$ and $i$.
Hence, $D$ lifts to an automorphism of the vector bundle $V$. As a consequence,
$V$ defines a symmetric vector bundle on the quotient space $Y$ 
with $N_{\rm gen}(Y)=3$. A symmetric vacuum can be constructed by taking, 
for example, $V_1=V_2=V$.
Finally, we must check the level matching constraint.
Having chosen $V_1=V_2=V$ then, for the first ${\bf Z}_3$ symmetry, 
we set $\tilde{k}_{1a}=\tilde{k}_{2a}=\tilde{k}_a$.
It is  easy to check that the level matching condition is
satisfied. Similarly, this can be verified for the second ${\bf Z}_3$.

To summarize, we have found a symmetric rank 3 vector bundle $V$ with
three generations. Choosing $V_1=V_2=V$, we obtain a symmetric vacuum
with low energy gauge group $E_6\times E_6$ and three
generations in the observable as well as in the hidden sector.
Furthermore, since the space $Y$ is not simply--connected, we can
introduce Wilson lines to break the observable sector gauge group $E_6$
spontaneously to $SU(3)\times SU(2)\times U(1)^3$.

\vspace{0.4cm}

Let us consider another example with similar properties, this time in
weighted projective space. We start with the intersection of two
polynomials of degree 6 in ${\bf CP}^5_{1,1,2,2,3,3}$, as in Example 5
above. From Table~\ref{weighted2}, we use the second to last bundle
\begin{equation}
 V=V(4,3,3,3;2,2,2,2,1,1,1,1,1)\; .
\end{equation}
This is a rank 5 bundle with $N_{\rm gen}=18$.
We consider the symmetry $D={\bf Z}_6$ generated by
\begin{equation}
 g:X^\n\rightarrow \a^{k_\n}X^\n
\label{kabob}
\end{equation}
with $\a =\exp (2\p i/6)$ and the charges $k_\n$ given by
\begin{equation}
 {\bf k}=(0,1,1,3,3,4)\; .
\end{equation}
One can choose two degree 6 polynomials that admit this symmetry and
define a non--singular manifold. The set of fixed points under the 
${\bf Z}_6$ transformations~\eqref{kabob} can be shown to be at most
a complex  curve. Again, generically, a three--fold $X$ in
${\bf CP}^5_{1,1,2,2,3,3}$ does not intersect a curve and, therefore,
the symmetry is freely acting. It follows that we can define the quotient
Calabi--Yau three--fold $Y=X/D$. We must now lift $D$ to an 
automorphism of the bundle $V$. To do this, we have to apply the same
procedure as in the previous example.  We choose ${\bf Z}_6$ charges
$\tilde{k}_a$ for the coordinates $\z_a$ and pick explicit polynomials
$Q_a^i(X)$. Then the combinations $\z_a Q_a^i(X)$ should be invariant
under $D$ for all $a$ and $i$. For the case at hand, let us choose
\begin{equation}
 \tilde{\bf k} = (1,1,0,0,4,4,5,5,0)\; .
\end{equation}
It is then easy to show that one can pick polynomials $Q_a^i(X)$ with
the correct properties. Hence $D$ lifts to an automorphism of the
bundle $V$. Furthermore, $V$ specifies a bundle on $Y$ with $N_{\rm gen}=3$.
A symmetric vacuum is obtained by choosing, for example,
$V_1=V_2=V$. Setting $\tilde{k}_{1a}=\tilde{k}_{2a}=\tilde{k}_a$ in
eq.~\eqref{lm1} and \eqref{lm2}, we can verify that the level matching
constraints are satisfied. 

In summary, we have found a symmetric rank 5 bundle with three
generations. The low energy theory has a
gauge group $SU(5)\times SU(5)$ with three generations in both the
observable and the hidden sector. Again, we can introduce Wilson
lines on $Y$ to break the observable sector gauge group
$SU(5)$ spontaneously to $SU(3)\times SU(2)\times U(1)$.

\section{Properties of symmetric vacua}

We would now like to discuss some of the properties of symmetric vacua
and their associated four-- and five--dimensional effective actions.

Let us begin with the implications for the four--dimensional effective
action. Generically, this action has two types of strong
coupling corrections at first non--trivial order. First, there is the
well--known threshold correction to the gauge kinetic
functions~\cite{hp1,low,hp2,losw2}
\begin{equation}
 f^{(1,2)} = S\pm \e_S\b_iT^i \label{gkf}
\end{equation}
where $f^{(1)}$ and $f^{(2)}$ correspond to the observable and hidden
sector. Secondly, there are corrections to the matter field K\"ahler
metric~\cite{low,losw2} which has the form
\begin{equation}
 Z_{IJ}=e^{-K_T/3}\left[K_{BIJ}-
        \frac{\e_S\b_i}{S+\bar{S}}\tilde{\G}^i_{BIJ}\right]\; .\label{km}
\end{equation}
Here $K_T$ is the K\"ahler potential of the T moduli, $K_B$ is a
bundle K\"ahler metric and $\tilde{\G}_B$ some associated connection. The
indices $I,J,\dots$ run over different generations. These
quantities have been defined in ref.~\cite{susy5} but need not concern
us in detail here. The corrections are of linear order in the strong
coupling expansion parameter $\e_S$ given by
\begin{equation}
  \e_S = \left(\frac{\k}{4\p}\right)^{2/3}
        \frac{2\p\r}{v^{2/3}}\; .\label{eps}
\end{equation}
Here, we recall that $\k$ is the 11--dimensional Newton constant and $\r$
is the radius of the orbifold while $v$ is the Calabi--Yau volume.
Furthermore, these corrections are proportional to the topological
charges $\b_i$. This observation is crucial in our context. As already 
mentioned, in terms of the underlying bundles these charges are specified by
\begin{equation}
\b_i(*\o^i)= c_2(V_1)-\frac{1}{2}c_2(TX)=-c_2(V_2)+\frac{1}{2}c_2(TX) \; .
\end{equation}
where $\{\o_i \}_{i=1,\dots ,h^{1,1}}$ is a basis of $H^{1,1}(X)$.
The quantities on the right  hand side of this equation are exactly
those that are set to zero for symmetric vacua. Hence we conclude that
all charges $\b_i$ vanish for symmetric vacua. Most importantly, this
implies the vanishing of the strong coupling corrections in the gauge
kinetic functions~\eqref{gkf} and the matter field K\"ahler
metric~\eqref{km} above. From the above argument, those corrections vanish for
topological reasons and, hence, irrespectively of the specific values
of moduli. Furthermore, the gauge kinetic functions are not expected
to receive corrections at higher loop order. Therefore, they are
perturbatively uncorrected for symmetric vacua and are simply given by
\begin{equation}
 f^{(1,2)} = S\; .
\end{equation}
This equation holds in the weakly coupled limit as well. However, there
it is valid only approximatively since $|\e_S\b_iT^i|\ll|S|$ in this
region of moduli space. For symmetric vacua, the important difference
is that the threshold correction vanishes exactly throughout all of
(large radius) moduli space. Similarly, for symmetric vacua the K\"ahler
metric does not receive strong coupling corrections to this order and is 
given by
\begin{equation}
 Z_{IJ}=e^{-K_T/3}K_{BIJ} \; .\label{km1}
\end{equation}

\vspace{0.4cm}

Let us next discuss the five--dimensional effective
action~\cite{losw1,elpp,losw2} of heterotic M--theory. This
theory is a five--dimensional $\cN =1$ supergravity theory coupled to two
four--dimensional $\cN =1$ theories on the orbifold planes. 
The strong
coupling corrections manifest themselves in a gauging of the bulk
supergravity. Specifically, a certain $U(1)$ isometry associated with the 
three--form axion in the universal hypermultiplet
coset space $SU(2,1)/U(2)$ is gauged. The gauge connection is the
fixed linear combination $\b_i{\cal{A}}^i$, where the sum runs over the 
graviphoton in the supergravity multiplet
and the $h^{(1,1)}-1$ vector fields in the vector supermultiplets. The 
$\beta_{i}$ coefficients are the above topological charges. As usual, 
this gauging implies the existence of potential energy terms in the bulk 
supergravity theory proportional to the coefficients $\b_i\b_j$. This 
potential obstructs flat space from being a solution of the equations 
of motion. Instead, the ``ground state'' of the five--dimensional theory 
turns out to be a non--trivial BPS double three--brane given by
\begin{equation}
\begin{aligned}
ds_{5}^{2} &= a(y)^{2}dx^{\mu}dx^{\nu}\eta_{\mu\nu}+b(y)^{2}dy^{2} \; \\
V &= V(y) \;
\end{aligned}
\label{habob}
\end{equation}
where
\begin{equation}
\begin{aligned}
a &= k_{1}V^{\frac{1}{6}} \; \\
b &= k_{2}V^{\frac{2}{3}} \; \\
V &= \left( \frac{1}{6}d_{ijk}f^{i}f^{j}f^{k} \right)^{2} \;
\end{aligned}
\label{zabob}
\end{equation}
and
\begin{equation}
d_{ijk}f^{j}f^{k}=H_{i},  \qquad H_{i}=2\sqrt{2}k_{3}\beta_{i}|y|+c_{i}
\label{wabob}
\end{equation}
Here V is the dilaton field, $f^{i}$ are functions of $y$, $d_{ijk}$ are 
the intersection numbers of the Calabi--Yau three--fold, $k_{a}$ and 
$c_{i}$ are constants and the $\beta_{i}$ are the topological charges.
To obtain the 
four--dimensional effective theory one has to reduce on this non--trivial 
BPS domain wall.  
Now for symmetric vacua, as we have
seen above, $\b_i=0$. Hence the gauging, the gauge connection and the 
associated potential terms are
absent in five--dimensional effective theories based on such vacua.
In this case, the functions $H_{i}$ in~\eqref{wabob} become constants, 
as do the functions $f^{i}$. Hence, $a$, $b$ and $V$ in~\eqref{zabob} are 
constants and the BPS three--brane~\eqref{habob} degenerates to a solution
with flat space--time $S^1/Z_2\times M_4$ and a constant dilaton.

\vspace{0.4cm}

Finally, we would like to discuss some properties of symmetric vacua.
Let us first review the general situation for (not
necessarily symmetric) vacua. The eleven--dimensional solution describing
a vacuum is determined as an expansion around a pure Calabi--Yau
background~\cite{W}. Only the first non--trivial terms in this
expansion are known and they have been determined in
ref.~\cite{low,nse}. The size of these first order corrections
is controlled by $\e_S$, defined above, and
\begin{equation}
 \e_R = \frac{v^{1/6}}{\p\r}
\label{hope}
\end{equation}
More precisely, the first order corrections are given as an expansion
in harmonics on the Calabi--Yau three--fold. Hence, they have a massless
part corresponding to zero eigenvalue harmonics (zero modes of the
Calabi--Yau space) and a massive part corresponding to the non--zero
eigenvalue harmonics. These massless and massive parts are of order
$\e_S$ and $\e_S\e_R$, respectively. 

It is clear that, generically, this linearized solution is sensible only
as long as $\e_S\ll 1$ and $\e_S\e_R\ll 1$. If these constraints are
violated, higher order terms in the equations of motion (for example
quadratic terms in the eleven--dimensional Einstein equation) become
important and the linear
approximation breaks down. Also, beyond linear order one expects
(partially unknown) corrections of order $\k^{4/3}$ to the
eleven--dimensional action to become important. At any rate, if the
parameters~\eqref{eps} and~\eqref{hope} approach unity the 
linearized supersymmetric
background is invalidated and, at present, there is no ``all order''
version that could replace it. This remains true even if one had arranged
both gauge couplings on the orbifold planes to be perturbative.
As a result, supersymmetric vacua are known only in a restricted
portion of the moduli space.

Let us now discuss what happens for symmetric vacua. 
As we have already mentioned, the corrections to the Calabi--Yau
background are caused by the non--vanishing source terms in the
Bianchi identity~\eqref{G}. For symmetric vacua, each source term in
this Bianchi identity vanishes in cohomology. As a consequence, the
massless part of the corrections vanishes. Indeed, the
massless part is proportional to $\e_S\b_i$ with the charges $\b_i$ 
defined above. For a symmetric embedding, the source terms in this
Bianchi identity, although zero in cohomology, might not,
in general, vanish identically. Correspondingly, the massive part of
the corrections to the vacuum does not necessarily vanish for symmetric vacua.

Symmetric vacua remove the first obvious
obstruction to making $\e_S$ large. To see this let us assume that
$\e_S\e_R\sim \k^{2/3}/v^{1/2}$ still remains small so that we do not
need to worry about the massive part of the solution. Then there are
no sizeable linear corrections to the Calabi--Yau background (since
the massless part vanishes) and, at the same time, higher order terms in
the equations of motion remain small. We still have to worry about
unknown correction of order $\k^{4/3}$ to the action that we have not
taken into account. Those may reintroduce large corrections at the
quadratic order~\footnote{Recently, such corrections have been studied
in ref.~\cite{Li} in a toy model.}. We do not know whether or not
this happens but we may at least speculate that symmetric vacua
are special enough to prevent such higher order corrections
to occur. Then such vacua would allow one to access the part
of the moduli space with $\e_S\geq 1$.

\vspace{0.4cm}

{\bf Acknowledgments} 
A.~L. would like to thank Dieter L\"ust and the group at Humboldt
University Berlin for hospitality. A.~L.~is supported by the European Community
under contract No.~FMRXCT 960090. B.~A.~O.~is supported in part by 
DOE under contract No.~DE-AC02-76-ER-03071 and by a Senior 
Alexander von Humboldt Award.



\end{document}